\documentclass[conference]{IEEEtran}

\usepackage{color}

\usepackage{cite}
\usepackage[pdftex]{graphicx}
\usepackage[ruled,vlined]{algorithm2e}
\usepackage{algpseudocode}
\usepackage{mathtools, cuted}
\usepackage{amsmath}
\usepackage{amsthm}
\usepackage{bm}
\theoremstyle{definition}

\usepackage{multicol}
\usepackage{amsfonts}
\usepackage{xcolor}
\usepackage{subcaption}
\usepackage{caption}
\usepackage{caption}
\usepackage{subcaption}

\ifCLASSINFOpdf
\else
\fi
\begin{document}
%
\title{Machine Learning Decoder for \\ 5G NR PUCCH Format 0}


\author{\IEEEauthorblockN{Anil Kumar Yerrapragada\IEEEauthorrefmark{1}, Jeeva Keshav S\IEEEauthorrefmark{2}, Ankit Gautam\IEEEauthorrefmark{3}, Radha Krishna Ganti\IEEEauthorrefmark{4}}
\IEEEauthorblockA{Department of Electrical Engineering\\
Indian Institute of Technology
Madras \\ Chennai, India  600036\\
Email: \IEEEauthorrefmark{1}anilkumar@5gtbiitm.in,
        \IEEEauthorrefmark{2}ee22d404@smail.iitm.ac.in,
		\IEEEauthorrefmark{3}ee20m008@smail.iitm.ac.in,
		\IEEEauthorrefmark{4}rganti@ee.iitm.ac.in
}
}


%


\maketitle

\begin{abstract}
5G cellular systems depend on the timely exchange of feedback control information between the user equipment and the base station. Proper decoding of this control information is necessary to set up and sustain high throughput radio links. This paper makes the first attempt at using Machine Learning techniques to improve the decoding performance of the Physical Uplink Control Channel Format 0. We use fully connected neural networks to classify the received samples based on the uplink control information content embedded within them. The trained neural network, tested on real-time wireless captures, shows significant improvement in accuracy over conventional DFT-based decoders, even at low SNR. The obtained accuracy results also demonstrate conformance with 3GPP requirements. 
\end{abstract}


\IEEEpeerreviewmaketitle

\section{Introduction}
 A wide range of wireless applications supported by 5G fall into three different use cases~\cite{nikolich2017standards, popovski20185g}: Enhanced Mobile Broadband (eMBB), Ultra-Reliable Low Latency Communications (URLLC), and Massive Machine Type Communications (mMTC). To support these use cases, a fundamental building block of a 5G cellular system is the Physical Uplink Control Channel (PUCCH) that carries  the Uplink Control Information (UCI).  UCI is fed back from the User Equipment (UE) to the base station (gNodeB) and contains information such as (1) Hybrid Automatic Repeat Request (HARQ) Acknowledgements for prior downlink  transmissions, (2) Scheduling Requests (SR) for uplink resources and, (3) Channel State Information (CSI) reports containing channel quality metrics that enable link adaptation and downlink resource allocation.  Depending on the 5G use case, the UE can use one of five PUCCH formats~\cite{5g_bullets}. Formats 0 and 1 can carry small UCI payloads (1 or 2 HARQ bits and an SR). Formats 2, 3, and 4 can transmit much larger payloads (HARQ, SR, and CSI reports). Formats 0 and 2 occupy up to two symbols in the time domain, making them ideal for low latency applications (URLLC). Formats 1, 3, and 4 can occupy up to 14 symbols, making them suitable for applications requiring enhanced capacity and coverage (eMBB and mMTC). We note that while the 5G standards~\cite{3gpp_38_211, 3gpp_38_212, 3gpp_38_213} offer detailed descriptions of the transmitter steps for PUCCH signaling, the receiver implementation is largely left open.

Format 0 encodes the information in the phase of a waveform in the frequency domain.  In this paper\footnote{A preprint of this paper appears at https://arxiv.org/pdf/2209.07861.pdf}, we utilise a Machine Learning (ML) based receiver for the robust and accurate demodulation of UCI transmitted using PUCCH Format 0. We make the following contributions:
\begin{itemize}
    \item We  recast the problem as an ML classification problem. Then, taking inspiration from the success of neural networks in other domains, we apply similar approaches. To our knowledge, this is the first work that applies ML to design a PUCCH Format 0 receiver. 
    \item We validate the trained neural network on test data obtained from MATLAB simulations. In addition, to incorporate variations in channel states, we  test the network on real-time over-the-air data captured from a 3GPP compliant 5G testbed at IIT-Madras~\cite{5gtbiitm}. We also present comparisons with conventional Discrete Fourier Transform (DFT) based approaches. 
    \item Applying ML models to solve wireless communication problems involves several key considerations such as realistic dataset generation, processing of complex numbers, and determining the optimum training SNR. As we describe our approach, throughout the paper we also provide insights into these considerations.
\end{itemize}

\section{Background on PUCCH Format 0}
PUCCH Format 0 is used to transfer 1 or 2 HARQ acknowledgements and/or an SR~\cite{3gpp_38_213}. Format 0 signalling does not contain any pilot reference signals (Demodulation Reference Signals (DMRS)) nor does it use Quadrature Amplitude Modulation (QAM) to modulate data. Instead, it transmits cyclically shifted versions of a pre-defined base sequence. The UCI is encoded in the cyclic shifts of a base waveform.

\subsection{Format 0 Allocation and Sequence Generation} 
\begin{figure}[ht]
\centering
\includegraphics[width=0.42\textwidth]{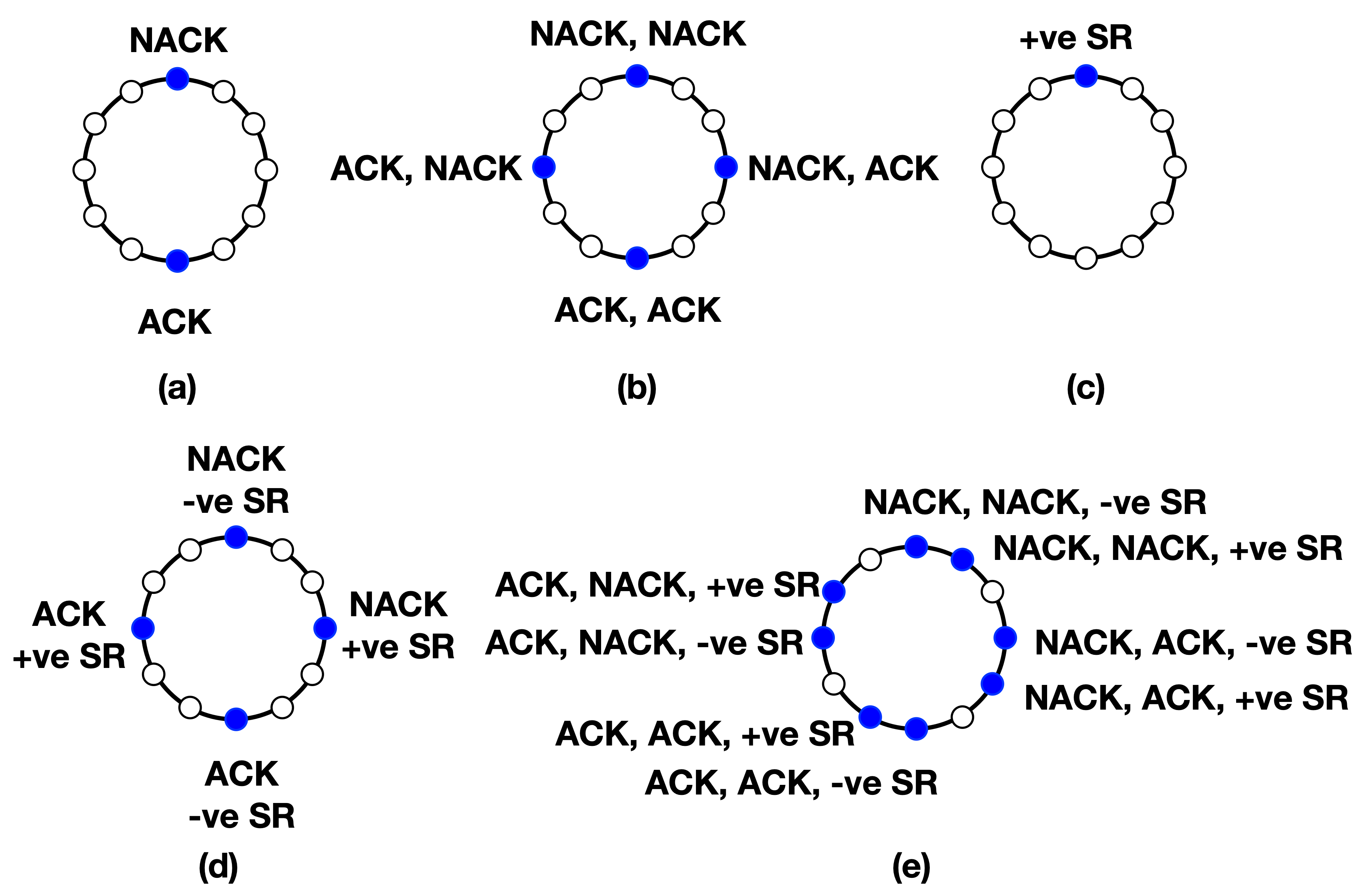}
\caption{The cyclic shifts $m_{cs}$ applied to the PUCCH Format 0 base sequence depend on the specific UCI content.}
\label{fig: m_cs table}
\end{figure}

The cyclically shifted base sequence is then mapped to the Resource Grid where it occupies one Resource Block (RB) in the frequency domain and either one or two symbols in the time domain. Overall, either 12 or 24 Resource Elements can be occupied. The second symbol holds a sequence similar to that of the first symbol and can be used for enhancement of SNR at the receiver. The PUCCH Format 0 sequence in the frequency domain in given by, 
\begin{equation}
        r_{u,v}^{(\alpha,\delta)}(n) = e^{j\alpha n}\cdot\bar{r}_{u,v} (n) = e^{j\alpha n}\cdot e^{j\phi (n) \pi /4} 
\end{equation}
where $n = 0,1,2,...,N_{sc}^{RB}-1$, where $N_{sc}^{RB}$ denotes the number of subcarriers per resource block. 
The sequence, given by $\bar{r}_{u,v}$, is the low Peak-to-Average Power Ratio (PAPR) base sequence~\cite{3gpp_38_211}. Lastly, $\phi(n)$ is given by Table 5.2.2.2-2 in~\cite{3gpp_38_211}. The subscripts $u$ and $v$ represent the group number and the sequence number within the group respectively. They are determined by higher layer group hopping parameters. We note that when PUCCH Format 0 is two symbols long, intra slot frequency hopping can be enabled. In this paper, for ease of exposition we assume intra slot frequency hopping  is disabled. 
The cyclic shift applied to the base sequence is denoted by $\alpha$ and is given by,
\begin{equation}
    \alpha = \frac{2\pi}{N_{sc}^{RB}}\bigg((m_{0} + m_{cs} + n_{cs}(n_{s,f}^{\mu},l+l'))\hspace{-8pt}\mod N_{sc}^{RB} \bigg)
    \label{eq: alpha}
\end{equation}

where,
\begin{itemize}
    \item $m_{0}$ is the initial cyclic shift. Note that Fig.~\ref{fig: m_cs table} is with respect to $m_{0}$ = 0. Other values of $m_{0}$ allow multiple users to be multiplexed on the same PUCCH Format 0 resources.
    \item $m_{cs}$ is the UCI specific cyclic shift, shown in Fig.~\ref{fig: m_cs table}. One bit of HARQ can be transmitted using cyclic shifts 0 and 6. Two bits of HARQ and can be transmitted using cyclic shifts 0, 3, 6 or 9. A single SR bit can be transmitted with a single cyclic shift of 0. A combination of one bit of HARQ and an SR can be transmitted using cyclic shifts of 0, 3, 6 or 9. A combination of two bits of HARQ and an SR can be transmitted using cyclic shifts of 0, 1, 3, 4, 6, 7, 9 or 10.
    \item $n_{cs}(n_{s,f}^{\mu},l+l')$ is a function based on the pseudo-random binary sequence defined in~\cite{3gpp_38_211}.
    \item $n_{s,f}^{\mu}$ is the slot number in the radio frame.
    \item $l$ is the OFDM symbol number in the PUCCH transmission where $l=0$ corresponds to the first OFDM symbol of the PUCCH transmission~\cite{3gpp_38_211, 3gpp_38_213}.
    \item $l'$ is the index of the OFDM symbol in the slot that corresponds to the first OFDM symbol of the PUCCH transmission in the slot~\cite{3gpp_38_211, 3gpp_38_213}.
\end{itemize} 

\subsection{Existing PUCCH Format 0 Receiver Methods}
Decoding the UCI format 0 involves detecting the UCI specific cyclic shift $m_{cs}$ applied to the base sequence by the UE. Format 0 decoding is different from other formats~\cite{kundu2018physical}. Since format 0 has no provision for DMRS, there is no channel estimation or equalization. Consequently, correlation based methods are used to decode the UCI. 

The low PAPR base sequence is known at the receiver. The decoding method involves correlating the received samples with various cyclically shifted versions of the base sequence. The applied cyclic shift is the cyclic shift that gives the highest correlation magnitude. The works in~\cite{kim2020performance, phan2021enhanced, tadavarty2021performance} provide a comparison of the correlation based receivers for various scenarios involving multiple antennas and multiple hops.

In this paper, we use the DFT based correlation for comparison. This method recovers the phase rotation $\alpha$ by taking the DFT of $e^{j \alpha n}\cdot \bar{r}_{u,v}(n)\cdot \bar{r}^{*}_{u,v} (n)$. The multiplication of the base sequence with its complex conjugate forces it to unity. The 12 point DFT then results in a peak at $\alpha$. Recall from~\eqref{eq: alpha} that $\alpha$ is a combination of $m_{0}$, $m_{cs}$ and $n_{cs}$. Therefore, on subtraction of $m_{0} + n_{cs}$ from $\alpha$ (subtraction is modulo 12), the UCI specific cyclic shift $m_{cs}$ remains. 

The DFT algorithm proves to be better from a hardware perspective due to its optimized use of resources (avoiding the need to correlate with all the shifted base sequences), reduced latency, and higher throughput. In the 5G testbed at IIT-Madras~\cite{5gtbiitm}, a DFT based receiver for PUCCH Format 0 has been implemented on custom  Field Programmable Gate Array (FPGA) boards.

\section{ML based receiver for PUCCH Format 0}
In this Section, we pose the detection of the cyclic shift $m_{cs}$ as a machine learning classification problem. Classification is a supervised learning task that involves predicting a label given an input data instance.  A classifier requires a training dataset, preferably with many instances of inputs and output labels from which to learn. A machine learning model such as a neural network can then look at the instances in the training dataset and learn the optimal mapping between input and output. It learns the mapping by minimizing a loss function. The loss function is some form of distance between the ground truth label and the label predicted by the neural network.  ML models are attractive because they can replace manual feature extraction rules and the need to have an explicit mathematical function mapping the input to the output. A well-trained neural network can extract highly complex non-linear features and mappings from a dataset that linear methods such as correlation may not capture. The capability of neural networks is evidenced by the fact that they have become adept at solving classification tasks across various domains, from computer vision to healthcare. 
\begin{figure}[ht]
\centering
\includegraphics[width=0.45\textwidth]{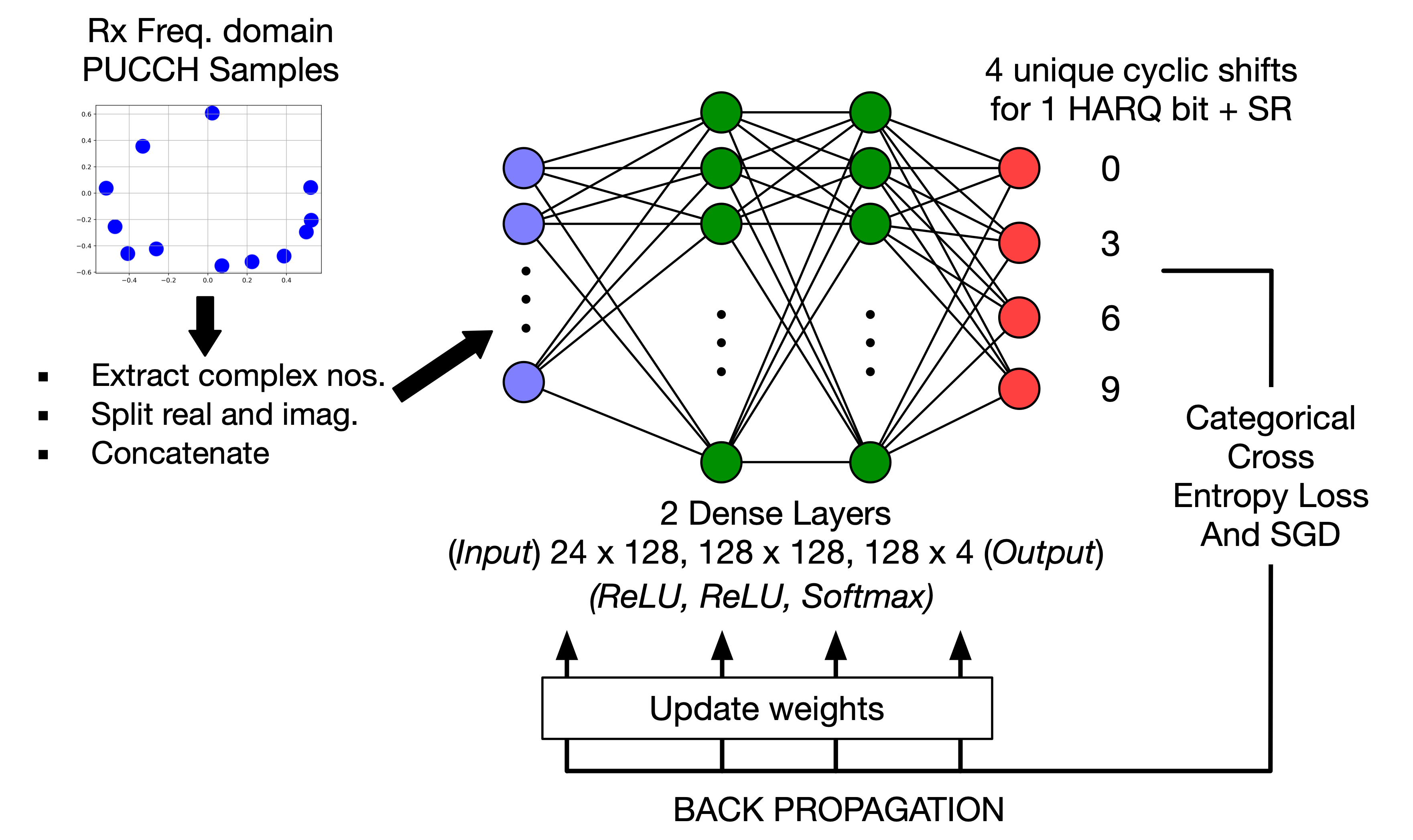}
\caption{PUCCH Format 0 neural network decoder architecture.}
\label{fig: pucch_nn_framework}
\end{figure}

\subsection{PUCCH Format 0 detection as a classification problem}
The classification task for the PUCCH Format 0 decoder is the following: Predict the applied cyclic shift $m_{cs}$ based on the received Format 0 signal samples (in the frequency domain). The predicted $m_{cs}$ in turn, maps to the UCI content (Fig.~\ref{fig: m_cs table}). The problem we are solving is very similar to image classification in that the received IQ samples are analogous to pixel values. Hence an equivalent neural network approach is expected to work.

After the cyclic prefix removal and the FFT of the time domain waveform at the receiver, we extract the resource block containing the 12 PUCCH Format 0 samples. These samples are the input to the neural network, as shown in Fig.~\ref{fig: pucch_nn_framework}. The output labels are the applied cyclic shift $m_{cs}$, which have to be known for each input instance during training. 

Note that the 12 resource block samples are complex numbers, leading to a key design issue. Commonly available neural networks and activation functions do not directly support complex number operations. Two options emerge. The first and most simple approach is to split each complex number into its real and imaginary components. Then either concatenate or interleave them to form one real vector of twice the size of the original complex vector. Another approach is to explore the notion of holomorphic neural networks and activation functions~\cite{bassey2021survey}. There is no proven theory that suggests that one method is better than the other. This paper chooses to adopt the first approach due to the ease of neural network implementation. Building real-valued neural networks is straightforward using commonly available Machine Learning libraries such as Tensorflow and PyTorch. 

\subsection{Dataset generation}
Dataset generation in ML for communication has a unique set of challenges. Firstly, benchmark datasets are not as widely available as in the image classification domain. Secondly, when seeking a model that one can deploy in a real-world system, it is desirable to use signal datasets captured from the actual hardware. Developing such datasets requires access to communication system hardware or testbeds. In the absence of such hardware, several state-of-the-art simulation tools such as the MATLAB 5G Toolbox exist to generate near-accurate datasets under various channel impairments. Furthermore, it is less tedious to create and pre-process large datasets using simulation tools than on hardware testbeds. 
Simulated data is a good starting point for training neural networks in communication problems. Hardware data, if available, can then be used to validate the performance in real world scenarios. For this paper, we use a combination of the two approaches. 

\subsubsection{Simulated data for training} Using the MATLAB 5G Toolbox, we generate received waveforms containing the PUCCH Format 0 signals. These waveforms include fading channel impairments and Additive White Gaussian Noise (AWGN). For various SNR values ($0$, $5$, $10$, $15$  and $20$ dB), we generate PUCCH signals transmitted over a TDL-C Channel and store the noisy received samples. These samples are the input data. For each input, the corresponding output label is the applied cyclic shift $m_{cs}$. 

\subsubsection{Hardware captures for testing}
\begin{figure}[ht]
\centering
\includegraphics[width=0.35\textwidth]{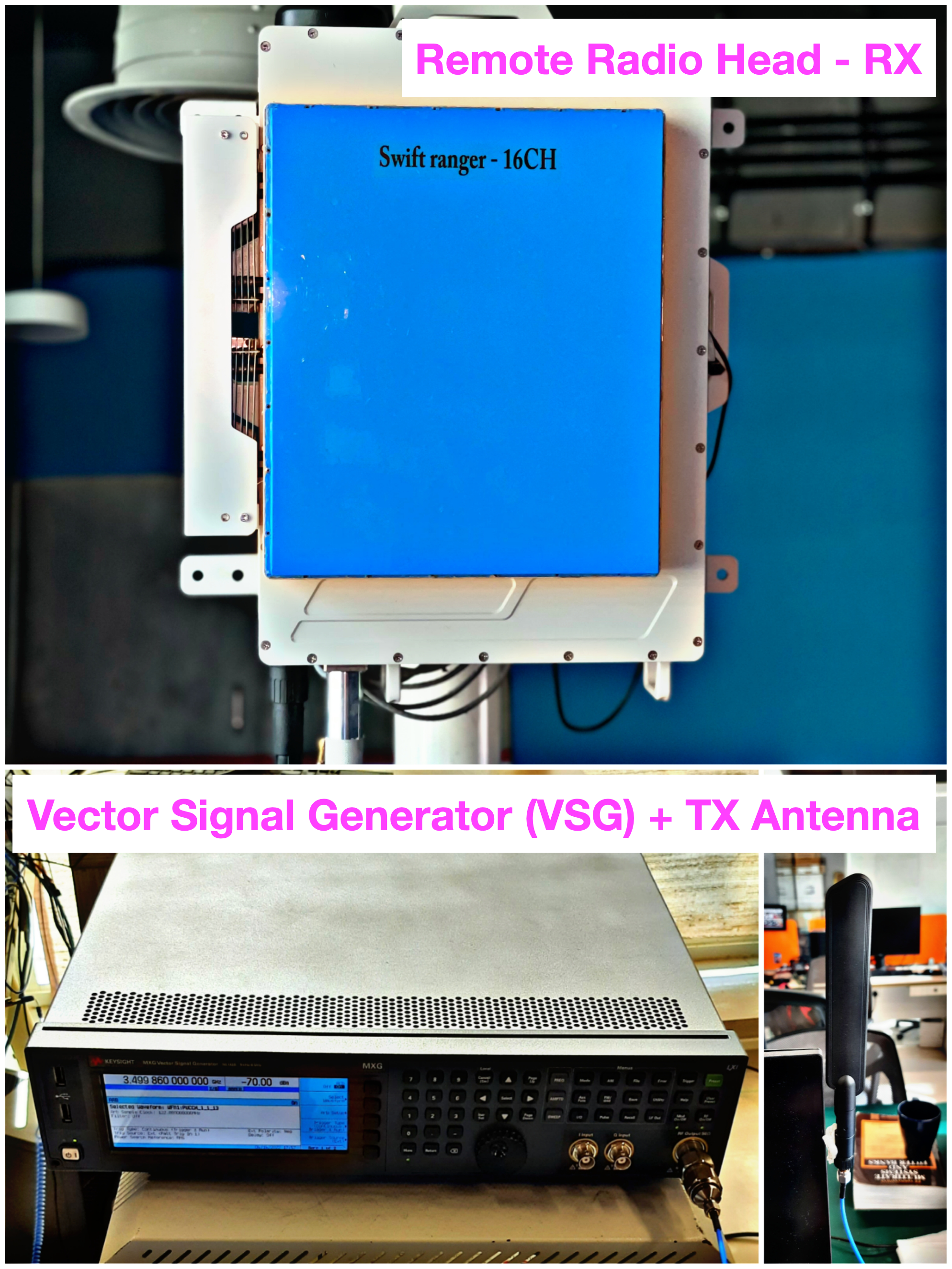}
\caption{IIT-Madras 5G testbed setup with the Remote Radio Head used as a receiver and the VSG used as a transmitter.}
\label{fig: hw_setup}
\end{figure}

We use hardware captures derived from the state-of-the-art 5G testbed at IIT Madras~\cite{5gtbiitm} for more realistic testing. The hardware captures help represent a wide range of channel states not included in the training dataset. The setup (see Fig.~\ref{fig: hw_setup}) consists of an N5182B Vector Signal Generator (VSG) for transmitting the  5G signal at a center frequency of $3.49986$ GHz (sub-6 GHz raster in the n78 band). The antenna used is a commercial omnidirectional wideband monopole. The antenna transmits signals originating from the Vector Signal Generator (VSG). The VSG is connected to the antenna through 2 SMA cables with $1.9$ dB wire loss each. 

A  multi-channel receiver front end connected to a dual-polarized antenna receives the signals from over the air (For the purpose of the paper, we utilise only one antenna and one transceiver chain). Other receiver components include an in-house Low Noise Amplifier (LNA) with 60dB gain at the receiver front end and an ADRV 9009 RF transceiver. We place the transmitter one meter away from the receiver to emulate a real-time wireless scenario. The PUCCH Format 0 signal is transmitted from the VSG through an antenna, over the air, then received at the LNA, followed by the transcevier. The signal out of the 16-bit ADC is then collected and used for testing. 

In both the software and hardware datasets, without loss of generality, we focus on the case where the UCI contains one HARQ bit and an SR. We generate 200000 instances for various SNR values. In each instance, the transmission includes a randomly chosen combination of 1 HARQ and SR bits and the corresponding cyclic shift $m_{cs}$ is applied to the base sequence. Our experiments revealed that the function $n_{cs}$ in~\eqref{eq: alpha}, which varies with slot and symbol index, is one of the characteristics of the data that neural network latched onto. Further, we conjecture that the neural network might be affected by changes in the initial cyclic shift $m_{0}$ as well. Since incorporating all combinations of $n_{cs}$ and $m_{0}$ into the training process is time prohibitive, in this paper we constrain ourselves to one of the commonly scheduled scenarios i.e., $m_{0} = 0$, symbol 13 and slots 13 and 14. 

\subsection{Neural Network Architecture}
There is no get-rich-quickly approach to determining the optimum neural network architecture. In the absence of such a formula, we have performed several experiments. Fig.~\ref{fig: pucch_nn_framework} shows one of the architectures that showed the best performance during training and testing. All performance curves in this paper are with respect to this configuration. Fig.~\ref{fig: nn_perf_configs} compares all architectures that were tested.

\subsubsection{Structure of the neural network} 
We employ a fully connected neural network with an input layer size of 24 neurons. The 24 corresponds to the concatenated real and imaginary parts of the 12 format 0 samples. The two intermediate layers are dense layers containing 128 neurons each. And the output neurons each represent one of four classes. 
\subsubsection{Regularization}
To prevent overfitting, we apply dropout between all the layers, with a probability of 0.5 (chosen by experiment). Dropout~\cite{srivastava2014dropout} is a regularization~\cite{Goodfellow-et-al-2016} technique that acts as a shortcut to achieve training the same model in multiple configurations of connections. It does so by randomly dropping some neurons during the training phase. A common cause for overfitting is neurons in subsequent layers compensating for "mistakes" made by those in prior layers. Dropout prevents this from happening, thus making training more noisy yet more robust at the same time. 
\subsubsection{Activations and Backpropagation}
Lastly, we add non-linearity to the neural network through Rectified Linear Unit (ReLU) activation functions in the dense layers. Since we require probabilities of each class at the final layer, we use the softmax activation function. Stochastic Gradient Descent (SGD) on the Categorical Cross-Entropy loss function makes up the backpropagation. In order to speed up learning while using SGD, we apply momentum which maintains an exponentially decaying moving average of prior gradients and moves in their direction~\cite{Goodfellow-et-al-2016}. 
\subsection{Training and Testing}
The choice of training SNR is far from trivial~\cite{o2017introduction}. Low SNRs obscure features of the data while high SNRs impede generalization against channel impairments. An impractical but straightforward approach would be to divide the data into SNR buckets and train a separate neural network for each bucket. 
However, ideally, one would like a single trained neural network to work over a wide range of SNRs. To this end, we train the neural network on 200000 data samples generated in MATLAB at a middle SNR of 10 dB. Note that we use 75\% of the dataset for training and the remaining for testing. We then test this neural network on simulated datasets of all 5 SNRs ($0$, $5$, $10$, $15$, and $20$ dB). Finally, we also use the same trained neural network to test with actual wireless hardware data captured using the IIT-Madras 5G testbed. We have settled on a training duration of 200 epochs 
and a learning rate of $10^{-3}$ through experiments. 
\begin{figure}[h!]
    \captionsetup{justification=centering}
     \centering
     \begin{subfigure}[b]{0.40\textwidth}
         \centering
         \includegraphics[width=\textwidth]{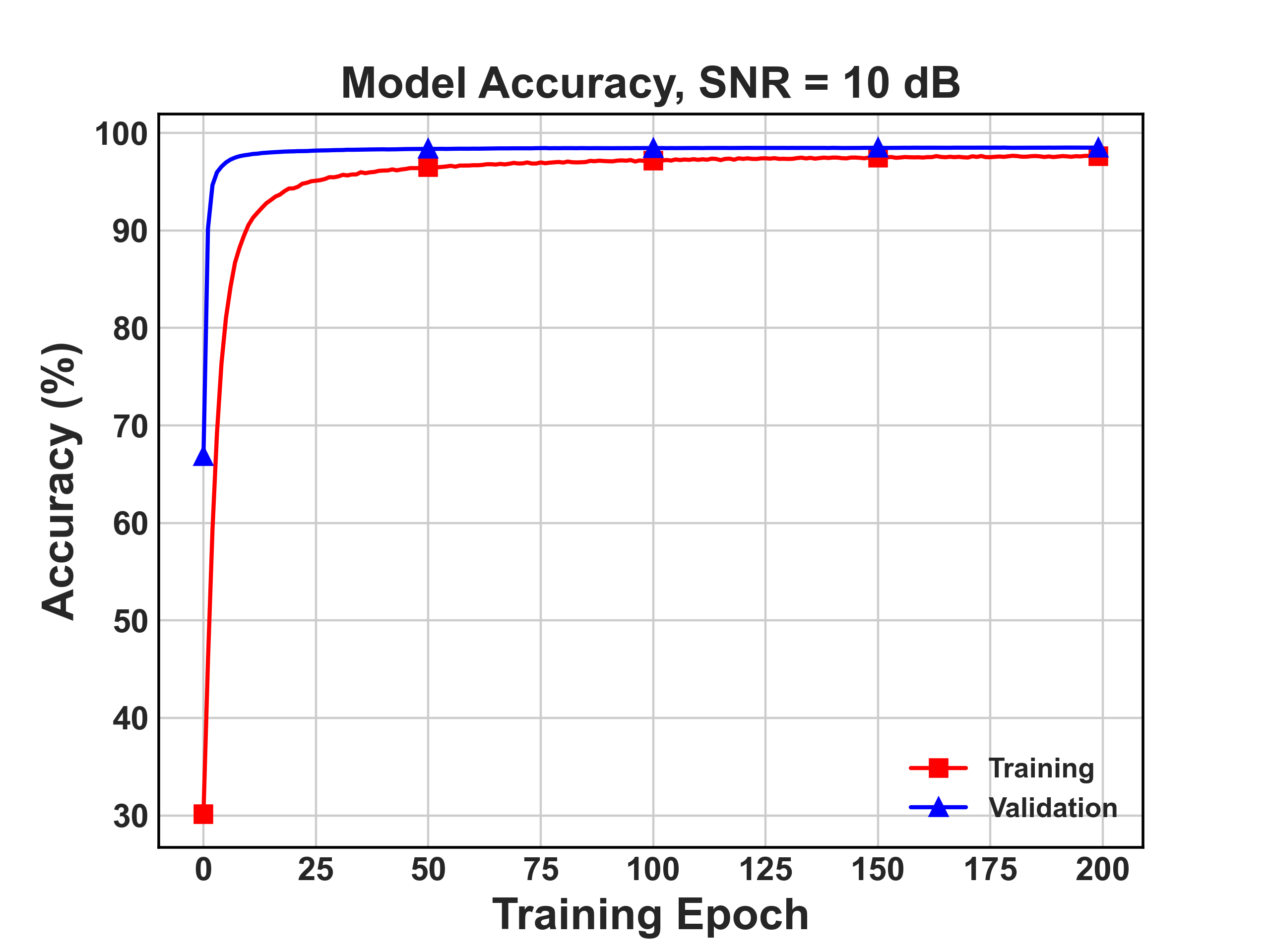}
         \caption{}
         \label{fig:accuracy_vs_epoch}
     \end{subfigure}
     \\
     \begin{subfigure}[b]{0.40\textwidth}
         \centering
         \includegraphics[width=\textwidth]{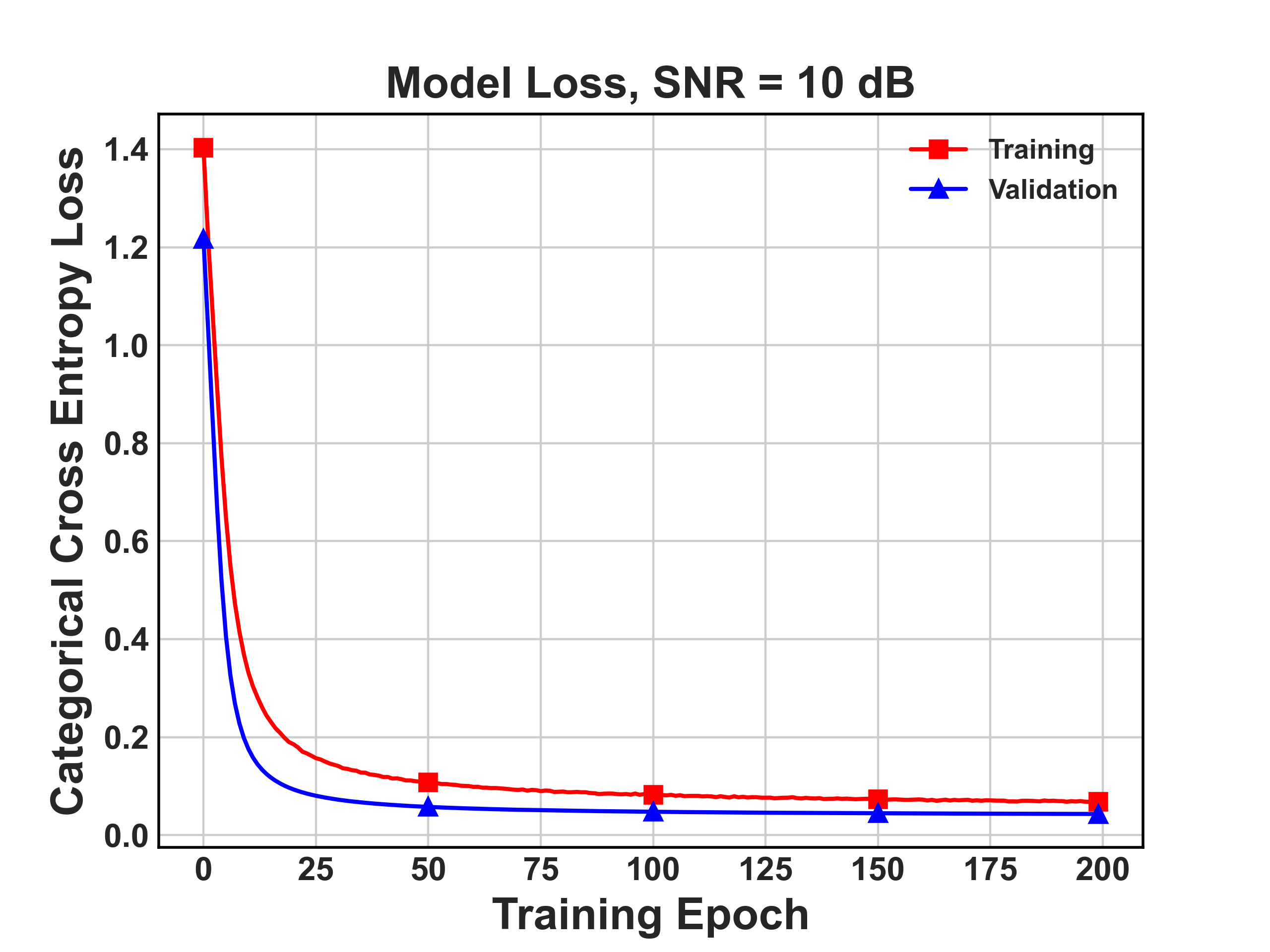}
         \caption{}
         \label{fig:loss_vs_epoch}
     \end{subfigure}
        \caption{(a) Training Accuracy and (b) Training Loss for simulated data at an SNR of 10dB}
        \label{fig:accuracy_loss_vs_epoch}
\end{figure}

\section{Results and discussion}
In this Section, we describe the results obtained from training and testing the neural network in various configurations. Prior work adopts two types of performance metrics~\cite{kim2020performance, phan2021enhanced, tadavarty2021performance} - False detection and Missed Detection. False detection refers to the gNB receiver predicting that UCI was sent when in fact nothing was sent. In this paper we confine missed detection to mean the gNB receiver incorrectly classifying a true UCI transmission by the UE, and not missing the detection. We adopt a slightly different metric (that also encapsulates missed detections), leaving false detection for immediate future work.

In keeping with other Machine Learning papers, we use accuracy as an evaluation metric. Accuracy is the ratio of correct predictions to the total number of predictions. It is the probability that the neural network believes a specific type of UCI was sent, given that some UCI was indeed sent by the UE. Furthermore, the accuracy metric by definition, incorporates the complement of missed detections - correct decoding of UCI inherently means that the gNB has not missed the detection. False detections can easily be incorporated into our framework by adding an additional class label whose inputs would be instances of AWGN. 

Fig.~\ref{fig:accuracy_loss_vs_epoch} shows the training and validation accuracy and loss. A generally increasing accuracy and decreasing loss, ultimately leading to convergence is indicative of a neural network that has learnt well. Furthermore, validation accuracy/loss values are slightly better, but not too much better than the training values suggesting that the neural network has not overfit. It also suggests that new unseen data has sufficient variance and is not skewed in favor of any one class. 

\begin{figure}[ht!]
    \captionsetup{justification=centering}
     \centering
     \begin{subfigure}[b]{0.4\textwidth}
         \centering
         \includegraphics[width=\textwidth]{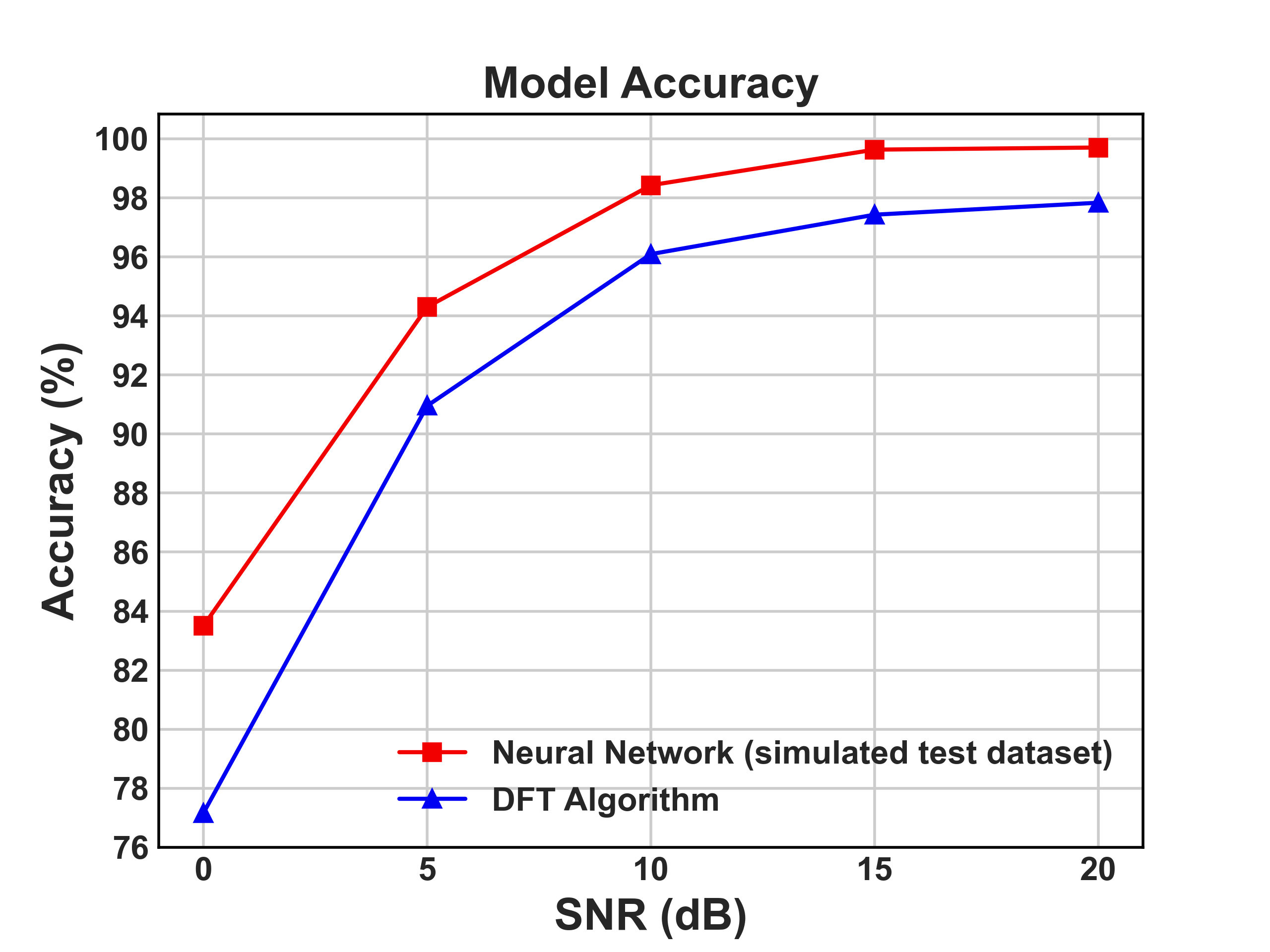}
         \caption{}
         \label{fig:accuracy_vs_snr_sim_nn}
     \end{subfigure}
     \\
     \begin{subfigure}[b]{0.4\textwidth}
         \centering
         \includegraphics[width=\textwidth]{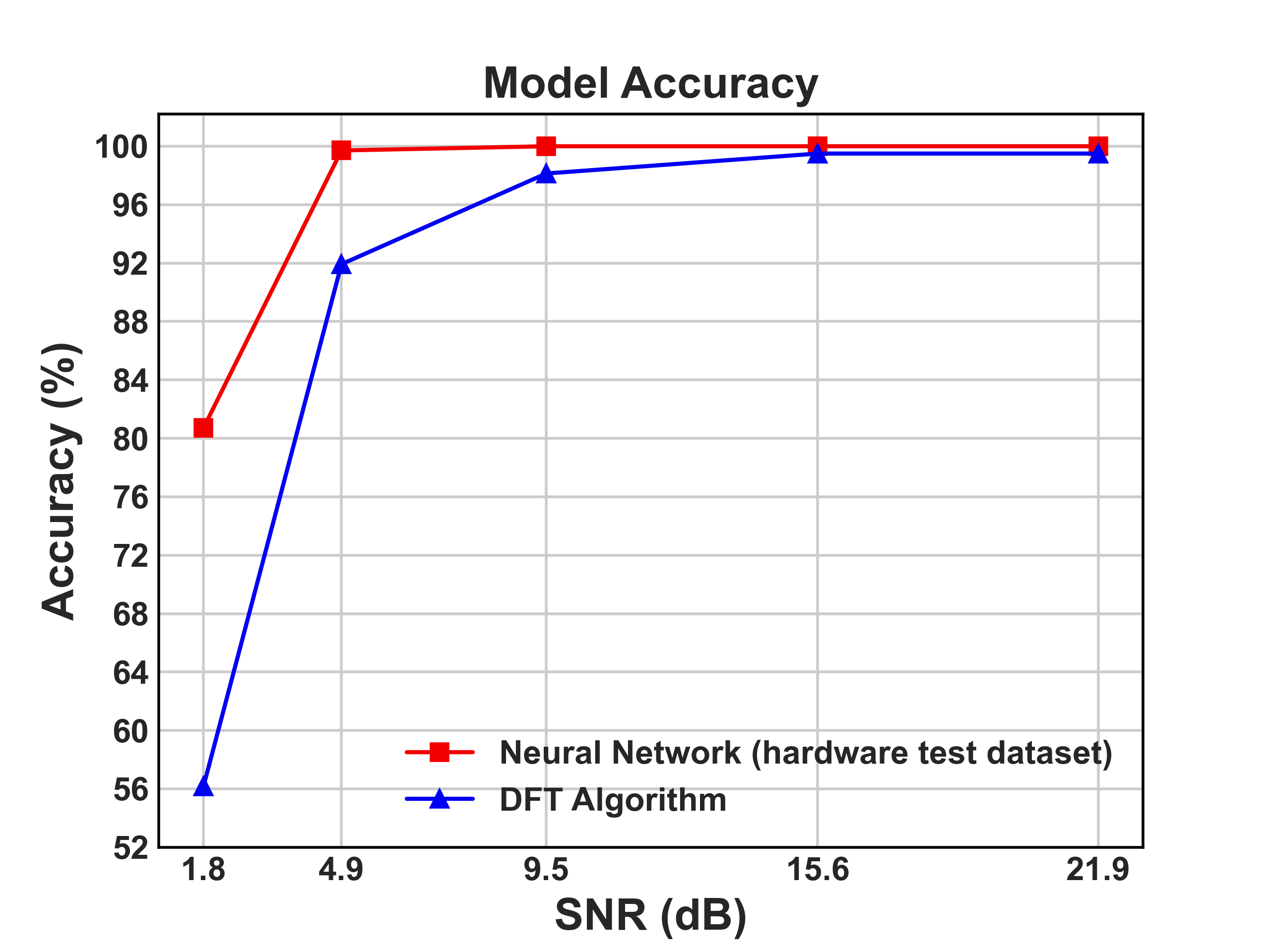}
         \caption{}
         \label{fig:accuracy_vs_snr_hw}
     \end{subfigure}
        \caption{Model accuracy on (a) simulated test data, (b) hardware captured test data. In both cases, accuracy is compared with that achieved by the DFT based decoder.}
        \label{fig: accuracy_vs_snr_sim_hw_fft}
\end{figure}

Test Accuracy as a function of SNR is shown in Fig~\ref{fig: accuracy_vs_snr_sim_hw_fft}\footnote{For SNR in Fig~\ref{fig: accuracy_vs_snr_sim_hw_fft}a, signal power is calculated considering the entire time domain received waveform for 1 slot. For SNR in Fig~\ref{fig: accuracy_vs_snr_sim_hw_fft}b, signal power is calculated by extracting the PUCCH location from the received spectrum.}. There is an increase in accuracy with SNR with simulated test data, as expected. The neural network shows a clear gain versus the DFT algorithm. Hardware test data follows a similar trend. Both simulations and hardware captures show accuracy values greater than 99\% for a wide range of SNR, thus surpassing the 3GPP conformance requirements~\cite{3gpp_38_141}. 

A significant gain in accuracy with the neural network is seen at the very low SNR end of the curves in Fig.~\ref{fig: accuracy_vs_snr_sim_hw_fft}. We observe about $3dB$ gain in performance of the neural network compared to the DFT method for simulated data. At higher SNRs, both methods seem to converge. This is more evident when the neural network is tested with the actual hardware samples. This might be because:
\begin{enumerate}
 \item As a result of the complex nature of setting up and operating hardware testbeds, there is currently a discrepancy in the sizes of the simulated and hardware datasets. At this time, the transfer of many hardware captures to the host computer is limited by the throughput of the Joint Test Action Group (JTAG) standard. Over time, we expect this to improve by adopting other methods. 
 \item The training data set (from MATLAB), might not incorporate all the non-idealities that are seen in real hardware. 
 \end{enumerate}

\begin{figure*}[t!]
    \captionsetup{justification=centering}
        \begin{subfigure}[b]{0.3\textwidth}
         \includegraphics[width=\textwidth]{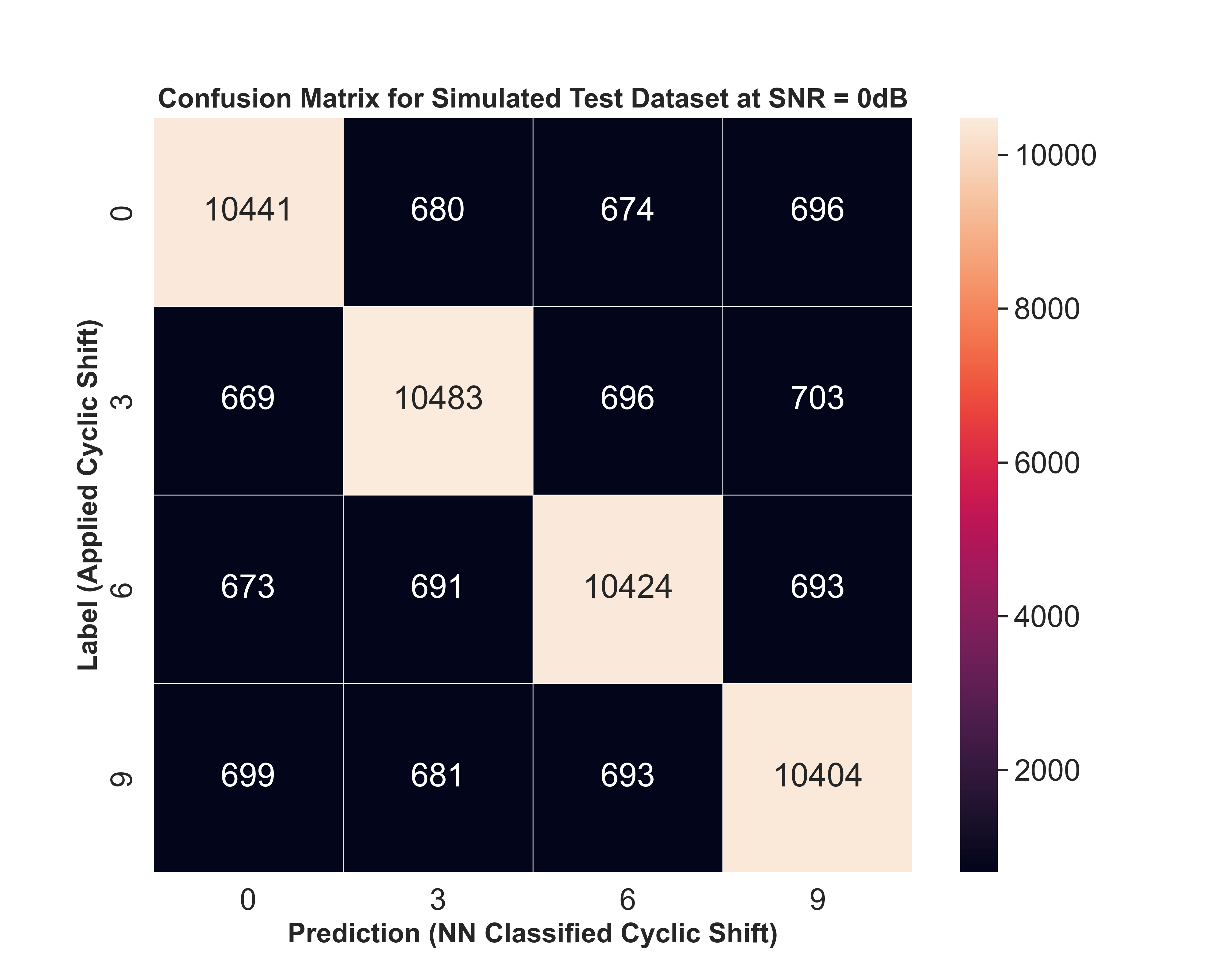}
         \caption{}
         \label{fig:conf_mtx_0_sim}
     \end{subfigure}
     \hfill
     \begin{subfigure}[b]{0.3\textwidth}
         \includegraphics[width=\textwidth]{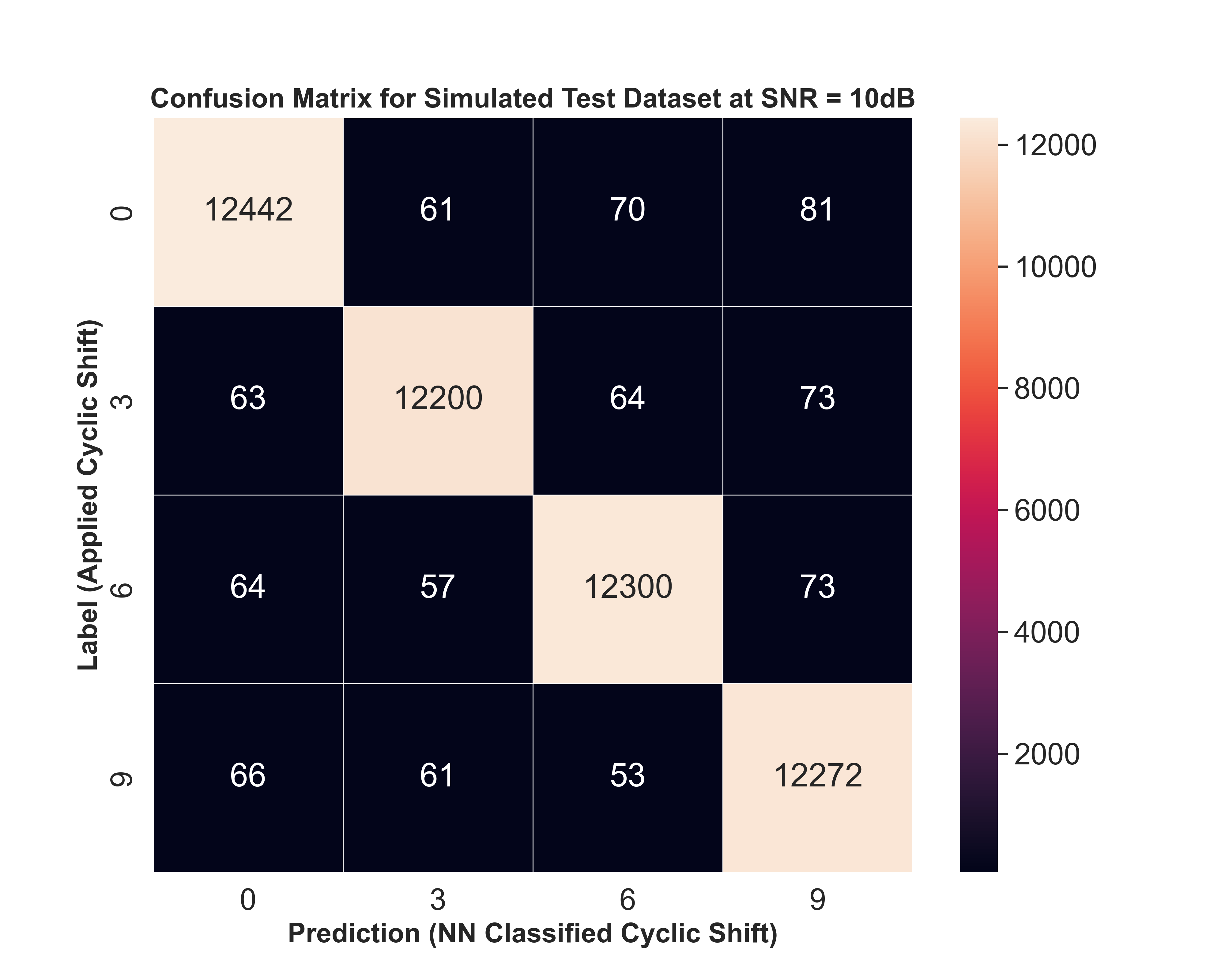}
         \caption{}
         \label{fig:conf_mtx_10_sim}
     \end{subfigure}
     \hfill
     \begin{subfigure}[b]{0.3\textwidth}
         \includegraphics[width=\textwidth]{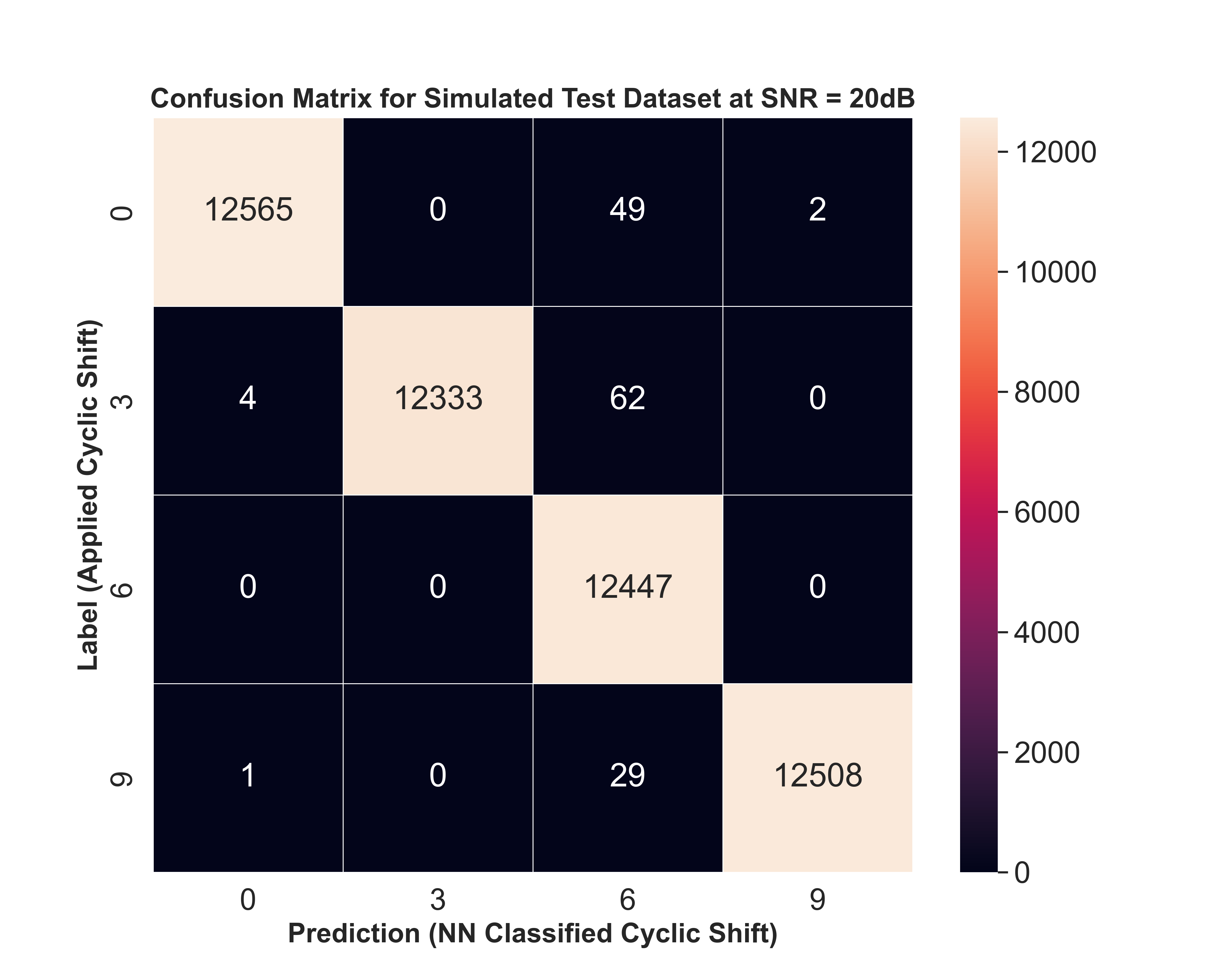}
         \caption{}
         \label{fig:conf_mtx_20_sim}
     \end{subfigure}
     \\
     \begin{subfigure}[b]{0.3\textwidth}
         \includegraphics[width=\textwidth]{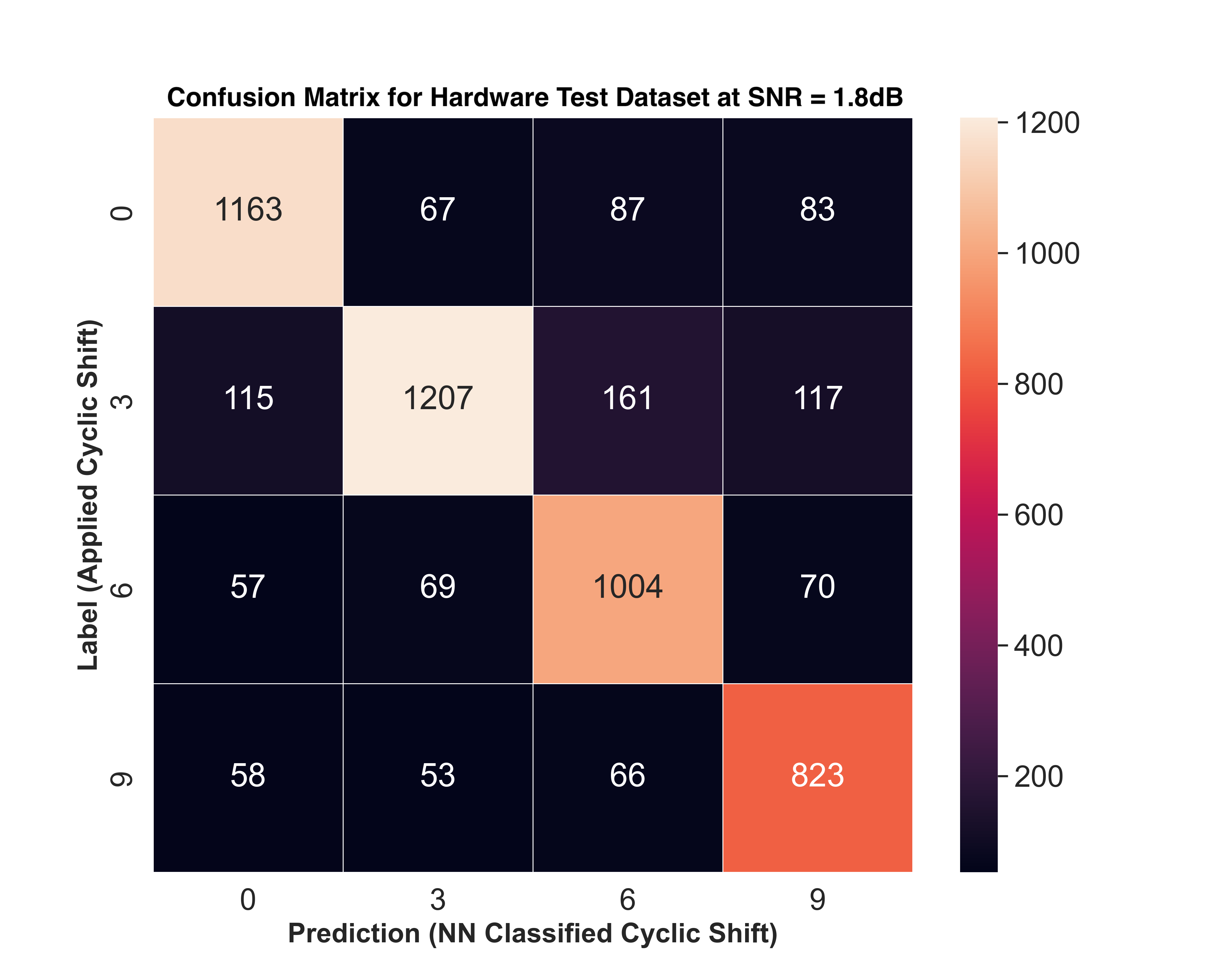}
         \caption{}
         \label{fig:conf_mtx_0_hw}
     \end{subfigure}
     \hfill
     \begin{subfigure}[b]{0.3\textwidth}
         \includegraphics[width=\textwidth]{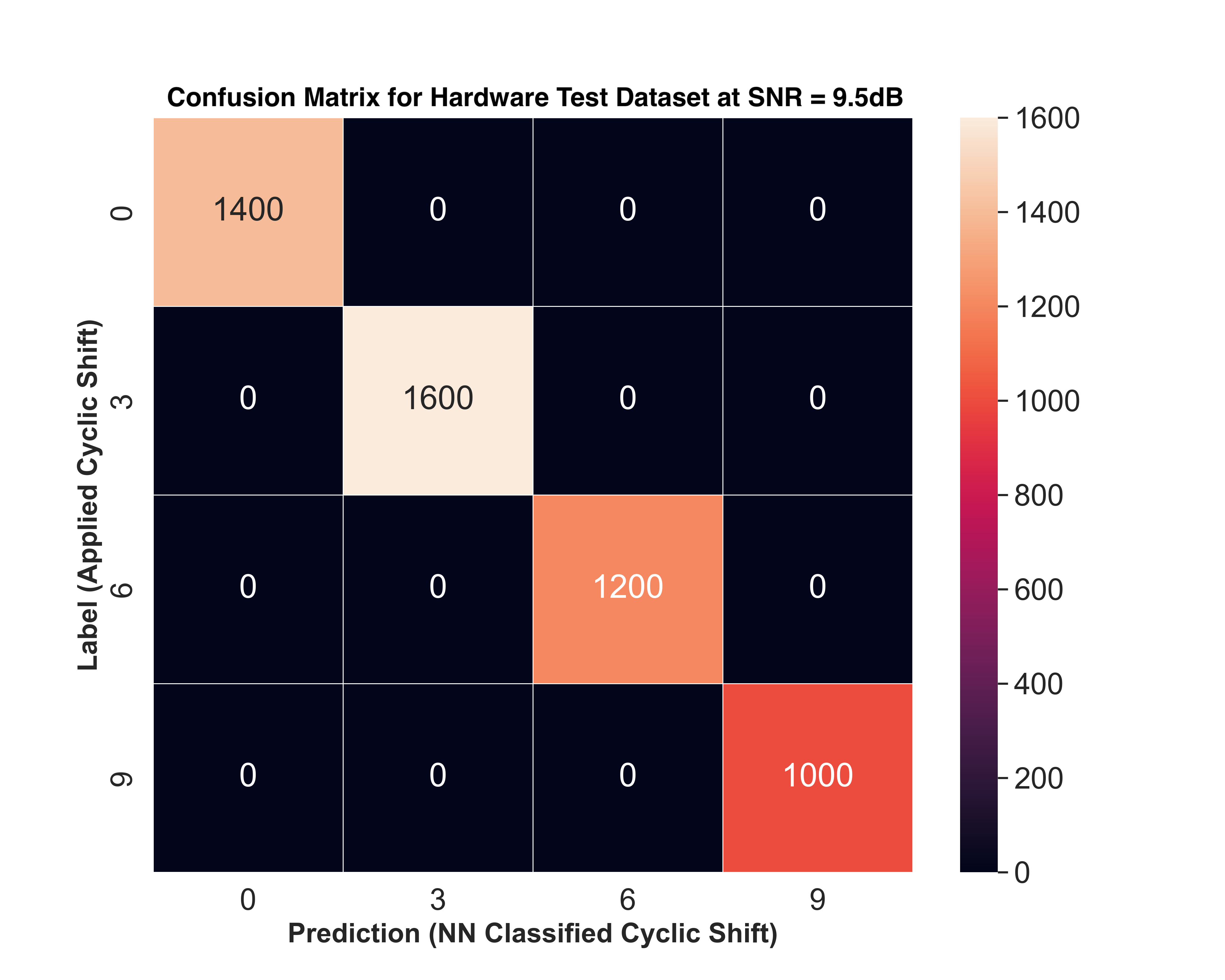}
         \caption{}
         \label{fig:conf_mtx_10_hw}
     \end{subfigure}
     \hfill
     \begin{subfigure}[b]{0.3\textwidth}
         \includegraphics[width=\textwidth]{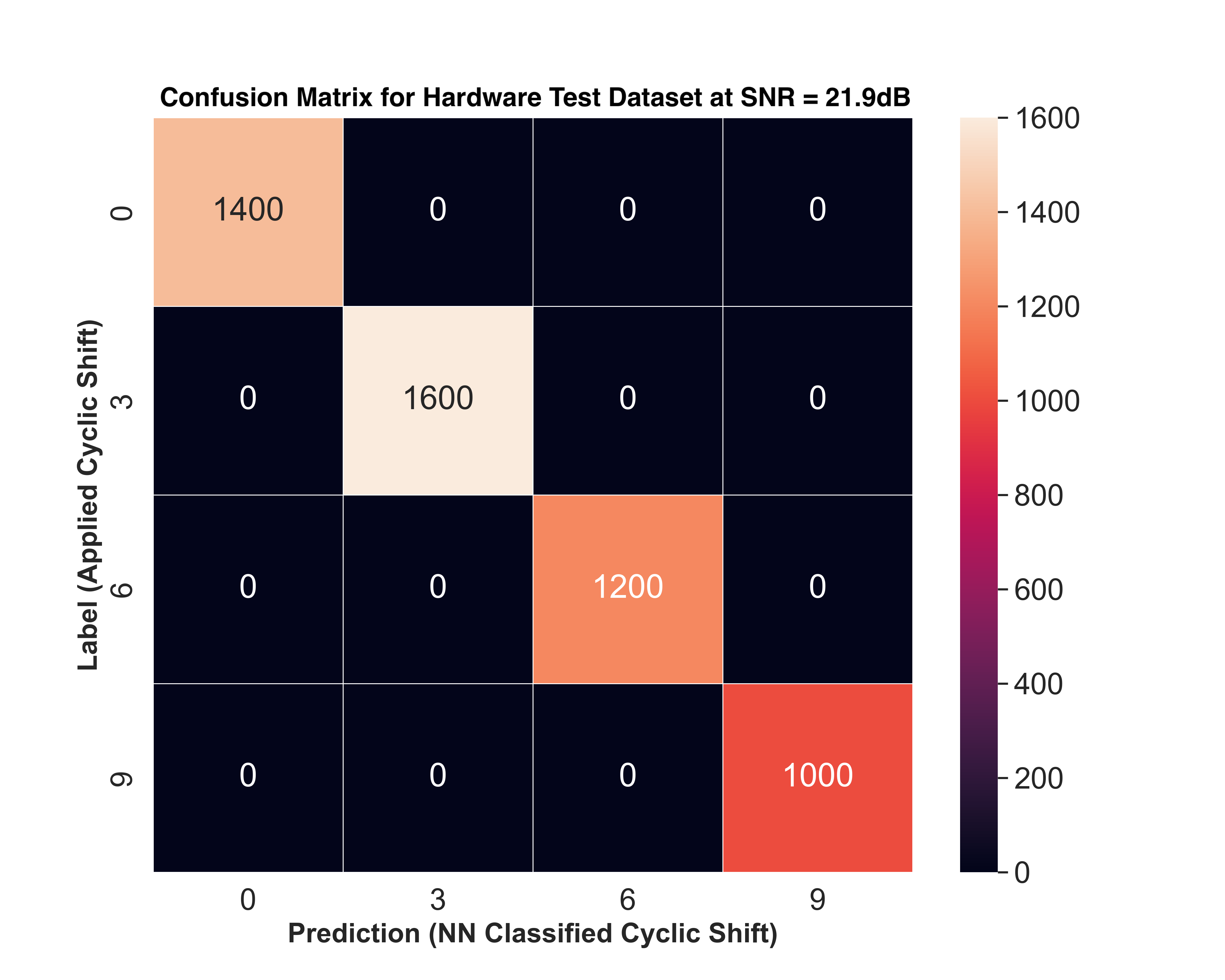}
         \caption{}
         \label{fig:conf_mtx_20_hw}
     \end{subfigure}
     
        \caption{Confusion Matrices for MATLAB simulated test data at (a) SNR = 0 dB (b) 10 dB (c) 20 dB. Confusion Matrices for hardware captured test data at (d) SNR = 1.8 dB (e) 9.5 dB (f) 21.9 dB }
        \label{fig:conf_mtx_sim_hw}
\end{figure*}

The confusion matrices for the neural network tested with both the simulated and hardware datasets in Fig.~\ref{fig:conf_mtx_sim_hw} show that the majority of the predictions made by the neural network are correct.

\begin{figure}[ht]
\centering
\includegraphics[width=0.45 \textwidth]{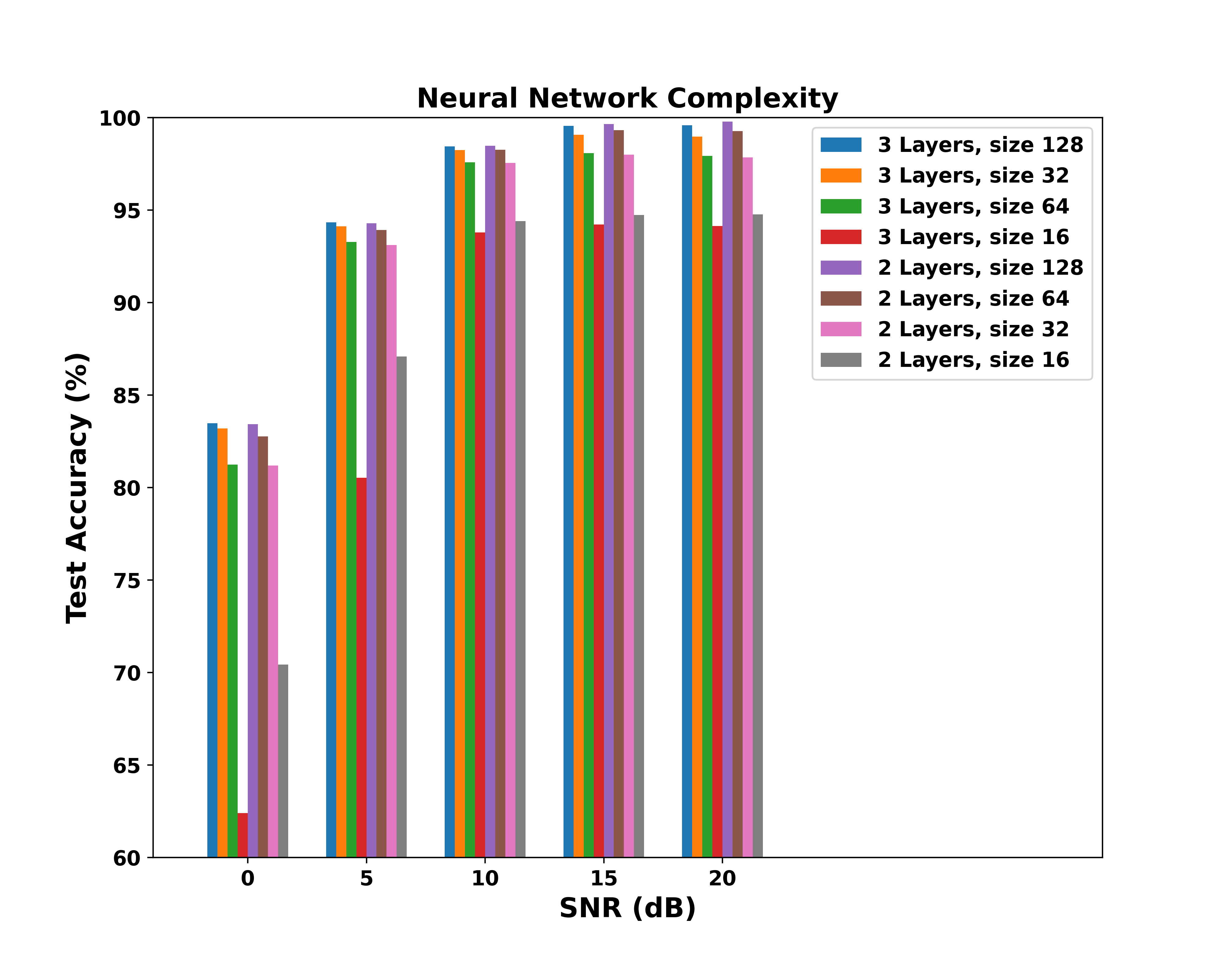}
\caption{Neural network performance for various complexities.}
\label{fig: nn_perf_configs}
\end{figure}

With the aim of making the neural network hardware-worthy, in this paper we begin an analysis on the impact of neural network size. Fig.~\ref{fig: nn_perf_configs} shows the test accuracy versus SNR for various configurations of the neural network. These experiments are performed on simulated test datasets. While in this paper, we have provided the results with a neural network with 2 layers and 128 neurons in each layer, we observe that similar performance can be obtained using 2 layers and 32 neurons in each layer, thereby significantly reducing the complexity of the network. 

In terms of complexity, the 2 layer 32 neuron model requires about 8 times and the 2 layer 128 neuron model requires about 96 times more multiplication units than the traditional DFT based implementation. This heavy complexity is outweighed by the improvement in performance of the receiver especially at lower SNRs. To fully quantify the trade-off between complexity and performance we note that more specific analysis with specific hardware devices and hardware captured datasets is necessary.

\section{Conclusion and Further Work}
In this paper, we have evaluated the performance of a neural network classifier in its ability to decode Uplink Control Information transmitted on the PUCCH using Format 0. Results show improvements in detection capabilities compared to mathematical approaches such as the DFT. 
Future work must investigate the feasibility of data generation and neural network training for all possible values of $m_{0}$, $m_{cs}$ and $n_{cs}$. Secondly, an extended classifier could include an additional class for false detections. Lastly, to aid real-time deployment of the trained neural network, it would be prudent to analyze its complexity from a hardware implementation perspective. 
\section*{Acknowledgment}
The authors would like to thank the Department of Telecommunications (DOT), India  for funding the 5G Testbed project and the Ministry of Electronics and Information Technology  (MeitY) for funding this work through the project "Next Generation Wireless Research and Standardization on 5G and Beyond".



\bibliographystyle{IEEEtran}
\bibliography{bibfile}

\end{document}